# Tunability of Excited-State Energy Levels of Four-Josephson-Junction Circuit in Crossover from Double-Well to Single-Well Potential

Yoshihiro SHIMAZU

*Department of Physics, Yokohama National University, Yokohama 240-8501*

A double-loop four-Josephson-junction (4-JJ) flux qubit is a tunable qubit wherein the potential profile can be varied *in situ* via two applied magnetic fluxes. This provides a high level of control over state transitions. Here, we report the results of spectroscopic measurements on this system. We observed microwave resonances associated with the first and second excited states. A branch of resonances that is most likely attributable to the third excited state was also observed. The excitation energy spectra for these levels were in qualitative agreement with the results of simplified calculations performed using an effective Hamiltonian with two phases. The three lowest levels characterized in this study may allow novel applications, including those proposed theoretically in the past. Our analysis showed that the potential exhibited a crossover from the commonly used double-well regime to the novel single-well regime. This indicates the possibility of manipulation and readout of a 3-JJ (and 4-JJ) qubit in the single-well regime using a simple technique.

1. **Introduction**

Nearly a decade of extensive research on superconducting qubits has shown that these are promising candidates for the implementation of scalable quantum information processors.[1] Among the various kinds of superconducting qubits, the flux qubit, which exploits the quantum coherence between different flux states, may have the longest coherence time due to the less sensitivity to background charge fluctuations. Since the first demonstration[2] of quantum superposition in a three-Josephson-junction (3-JJ) flux qubit,[3] many researchers have experimentally worked on 3-JJ qubits and its improved versions. A tunable coupling scheme[4] and controlled-NOT quantum gate operation[5] using a qubit pair have been demonstrated. The double-loop 4-JJ flux qubit is a variant of a 3-JJ flux qubit but has additional tunability.[6] Very recently, vacuum Rabi oscillations between an LC oscillator and a tunable 4-JJ flux qubit at a symmetry point have been observed.[7] A tunable RF superconducting quantum interference device (SQUID) comprising of a superconducting loop intersected by a small DC-SQUID is another type of flux qubit;[8] many experimental studies have been performed on the tunable RF-SQUID as well. A great advantage of the 3-JJ and 4-JJ flux qubits over a tunable RF-SQUID is that one does not require a significant loop inductance to realize the double-well potential that is needed for implementation of a flux qubit. Therefore, the loop size of the 3-JJ and 4-JJ flux qubits can be reduced considerably, thereby increasing their immunity to external noise. As a result, these qubits will have longer coherence times than a tunable RF-SQUID.

Recently, quantum dynamics involving the three lowest levels of the "3-JJ flux qubit" geometry have attracted much attention. Electromagnetically induced transparency[9] and microwave amplification have been demonstrated using this geometry,[10] which has been used as a flux qubit in earlier experiments. The system studied in ref. 10, a single-loop 4-JJ circuit, is effectively equivalent to the commonly studied 3-JJ flux qubit. Meanwhile, You *et al.* have proposed applications using the three lowest levels of a "double-loop 4-JJ flux qubit" geometry, that is, single-photon production[11] and simultaneous cooling[12] of a qubit and its neighboring quantum system. This system provides an enhanced level of controllability over the 3-JJ flux qubit. Both energy levels and transition matrix elements can be controlled via two magnetic fluxes that are independently controlled in the 4-JJ flux qubit. In particular, the transition matrix element between the two lowest levels can be tuned over a very wide range. This property is advantageous for realizing state population inversion and its inverse process.[11,12] However, in previous experiments conducted on this system, only the two lowest levels have been characterized[6] and utilized in a quantum optics-like experiment.[7]

We studied the level structure of the 4-JJ qubit using a simple spectroscopic technique.[2] We clearly observed one- and two-photon resonances associated with the transitions from the ground state to the first and second excited states. Resonances that were most likely caused by transitions to the third excited state were also observed. The observed level structure for various control fluxes were in qualitative agreement with the results of a simplified calculation performed using an effective two-phase Hamiltonian, which was also used in refs. 11 and 12. This result forms a solid basis for realizing applications that use the three lowest levels. We experimentally identified the operating point at which the three energy levels are equidistant. This operating point may allow novel quantum-information applications.

Our analysis indicates that the potential exhibited a crossover from the double-well regime to the single-well regime. Thus, we demonstrated the possibility for using a 3-JJ (and 4-JJ) qubit, not in the flux-qubit (double-well) regime but in the single-well regime, in the processes required for qubit implementation, namely, initialization, manipulation, and readout. The operation speed in the single-well regime can be faster than that in the commonly used double-well regime because of the large separation between the two lowest levels in the single-well regime. We note that a high level of quantum coherence associated with the single-well regime has very recently been demonstrated in ref. 13 for a 3-JJ qubit with a shunted capacitor.[14)]

## 2. Sample and Hamiltonian

A schematic representation and an optical microscopy photograph of the sample are presented in Fig. 1. The thick lines in the schematic denote the 4-JJ flux qubit. The magnetic flux produced by the circulating current in the main loop of the qubit is measured by an unshunted readout DC-SQUID, which is galvanically connected to the qubit main loop. A control line is inductively coupled to the DC-SQUID loop ($\alpha$-loop) incorporated in the qubit. The sample was fabricated using $e$-beam lithography with ELIONIX ELS-3300 and shadow deposition of Al on an oxidized Si substrate. An on-chip shunt capacitor for a readout DC-SQUID that is commonly used in experiments on flux qubits was not fabricated in this sample for the sake of simplicity of fabrication.

The potential energy profile of the 4-JJ qubit is manipulated through the magnetic frustrations $f_1$ and $f_2$ in the two loops, where $f_i = \Phi_i/\Phi_0$ ($i = 1, 2$) and $\Phi_0 = h/2e$. The areas of the qubit loops enclosing $f_1$ and $f_2$ are 74 μm$^2$ and 32 μm$^2$, respectively. $f_1$ and $f_2$ are controlled via a magnet current $I_{mag}$ flowing in a superconducting magnet and control current $I_{cont}$ according to the equations

$$f_1 = (M_1 I_{mag} + M_{c1} I_{cont} + \Phi_1^{(0)})/\Phi_0 \tag{1}$$

$$f_2 = (M_2 I_{mag} + M_{c2} I_{cont} + \Phi_2^{(0)})/\Phi_0, \tag{2}$$

where $M_i$ ($i = 1, 2$) are mutual inductances between the magnet line and qubit loops, $M_{ci}$ are mutual inductances between the control line and the loops, and $\Phi_i^{(0)}$ are static background magnetic fluxes. The mutual inductance $M_{SQ}$ between the readout SQUID loop and magnet line is determined from the period of the oscillations of the SQUID switching current as a function of $I_{mag}$. Both $M_1$ and $M_2$ are then derived from $M_{SQ}$, using the areas of the qubit loops and the SQUID loop. Because $M_{c1}$ is much smaller than $M_{c2}$, it can be neglected in the present analysis, and the variation in $I_{cont}$ leads to the variation in $f_2$, which in turn controls the minimum energy splitting (gap), whereas the variation in $I_{mag}$ significantly affects both $f_1$ and $f_2$. $M_{c2}$ is determined from the observed position of the qubit step as described below.

For a 3-JJ qubit and 4-JJ flux qubit as its variant, it is common to design a sample such that two of the junctions in the main loop have an equal area, in order to simplify theoretical treatment. For this specific sample under investigation, however, these junctions had different areas, $S = 0.028$ μm$^2$ and $\alpha_2 S$ with $\alpha_2 = 0.8$. This was not intentional but due to fabrication variations. Meanwhile, on the basis of the inspection of the sample, we assumed that the area of each junction in the α-loop was equal to β$S$, where β was approximately 0.6. It should be noted that a possible asymmetry in the α-loop could easily be taken into account in the calculation.[15]

The Hamiltonian, which is the sum of the potential energy term $U$ and kinetic energy term $K$, can be conveniently expressed as a function of $\varphi_p = (\varphi_1 + \varphi_2)/2$ and $\varphi_m = (\varphi_1 - \varphi_2)/2$, where $\varphi_1$ and $\varphi_2$ are the phase differences of the two junctions in the main loop with areas $S$ and $\alpha_2 S$, respectively. They are given by the expressions

$$U = E_J \left[ 1 + \alpha_2 + 2\beta - (1+\alpha_2)\cos\varphi_p \cos\varphi_m + (1-\alpha_2)\sin\varphi_p \sin\varphi_m \right. \\ \left. - 2\beta\cos(\pi f_2)\cos(2\pi(f_1 + f_2/2) + 2\varphi_m) \right] \quad (3)$$

and

$$K = \frac{E_c}{\alpha_2 + 2\beta(1+\alpha_2)}\left[ -(1+\alpha_2+8\beta)\partial_p^2 - (1+\alpha_2)\partial_m^2 + 2(1-\alpha_2)\partial_p\partial_m \right], \quad (4)$$

where $\partial_p = \partial/\partial\varphi_p$, $\partial_m = \partial/\partial\varphi_m$, $E_J = \hbar I_c/2e$, and $E_c = e^2/2C$. Here, $I_c$ and $C$ are the critical current and capacitance of the larger junction in the main loop, respectively. The above Hamiltonian was derived as a straightforward extension of that for a conventional 4-JJ qubit with two equivalent junctions ($\alpha_2 = 1$).[3] It should be noted that the loop inductances associated with the qubit loops are neglected in this effective two-phase Hamiltonian. The inductive effect will be discussed below.

In a conventional 3-JJ qubit, the barrier height of the potential strongly depends on the junction area ratio α.[3] In the system under investigation, the parameter

$$\alpha_{eff} = \frac{4\beta\cos(\pi f_2)}{1+\alpha_2} \quad (5)$$

plays the role of α in the conventional 3-JJ qubit. This value can be tuned *in situ* via $f_2$. When $\alpha_{eff} > 1/2$, the potential is a double well, whereas when $\alpha_{eff} < 1/2$, the potential is a single well. The gap increases exponentially with decreasing $\alpha_{eff}$ when $\alpha_{eff}$ is in the commonly used range, around 0.7–0.8.[3]

## 3. Experimental

Measurements were performed at the base temperature of a dilution refrigerator at approximately 20 mK. The sample was mounted in a copper cell, which was thermally anchored on the cold finger of a dilution refrigerator. A superconducting NbTi magnet (diameter 34 mm and height 6 mm) was located approximately 25 mm away from the sample. The principle of the measurement is the same as that in ref. 2. The switching current $I_{SW}$ of the readout DC-SQUID was measured to detect the state of the qubit. The SQUID bias current $I_{bias}$ was swept using an asymmetric triangular waveform, typically at 8 kHz, and the average switching current was recorded as a function of $I_{cont}$ and $I_{mag}$ using a sample-and-hold circuit and low-pass filter. $I_{cont}$ and $I_{bias}$ were supplied through twisted pairs of constantan leads. These lines were filtered by using a combination of two types of low-pass filters—lumped circuit filters and stainless steel powder filters. The voltage across the DC-SQUID was measured using the $I_{bias}$ line in a two-terminal measurement configuration. Magnetic shielding was provided using a superconducting lead shield and a permalloy shield. For spectroscopic measurements, microwaves were continuously applied using a small loop antenna at a distance of approximately 4 mm from the sample, rather than an on-chip microwave line, which is commonly used in experiments on superconducting qubits.

## 4. Results and Discussion

At a fixed value of $I_{cont}$, the dependence of $I_{SW}$ on $I_{mag}$ exhibited a clear qubit step at some values of $I_{mag}$, which indicated the reversal of the circulating current in the main loop of the qubit. The necessary condition for the appearance of the qubit step is $f_1 + f_2/2 = 0, \pm 1/2, \pm 1, \pm 3/2, \cdots$.[16] As $I_{cont}$ varied, the step position in terms of $I_{mag}$ changed linearly as a function of $I_{cont}$. From this dependence, the mutual inductance $M_{c2}$ was estimated to be 6.1 pH. For a specific qubit step, we observed strong dependence of the shape of the qubit step on $I_{cont}$, as shown in Fig. 2, indicating the flux signal measured with the readout SQUID as a function of $I_{mag}$ for various values of $I_{cont}$. The qubit step became smeared out with decreasing $I_{cont}$. This behavior should be closely associated with the variation in the energy gap.[16]

Excitation energy spectra associated with this specific qubit step for different values of $I_{cont}$ are shown in Figs. 3(a)–(c). These data were acquired in switching-current measurements under continuous microwaves. Each color plot shows a collection of peak and dip structures in the $I_{SW}$ vs. $I_{mag}$ curves for a constant microwave frequency $F$. These curves are obtained by subtracting the smooth background curve of $I_{SW}$ from the observed $I_{SW}$. The frequency $F$ was varied below 20 GHz. We see several branches of energy dispersion curves, including the commonly observed branches that correspond to one- and two-photon resonances associated with the transition to the first excited state. As shown clearly in Figs. 3(b) and 3(c), the energy dispersion curves as a function of $I_{mag}$ are asymmetric. This is because the control flux $f_2$ and, in turn, the barrier height changes with increasing $I_{mag}$. For these data, the barrier height increases with increasing $I_{mag}$. Therefore, the slope of the dispersion curve is steeper to the right of the minimal point.

To understand the origin of the observed branches of the excitation energy spectra, we compare the experimental data with the result of calculations that use the effective two-phase Hamiltonian [eqs. (3) and (4)]. The unknown parameters that enter the calculation are $E_J$, $E_c$, $\beta$, and $f_2^{(0)}$, where $f_2^{(0)}$ is the value of $f_2$ at the optimal point for $I_{cont} = 0$; the optimal point is the point at which the energy splitting between the ground state and the first excited state is minimal. By using the values given in the figure caption, a qualitative agreement between the data and the calculated results, shown in Figs. 3(d)–(f), is obtained. We note that the dependence of the energy levels on these parameters is very strong and that a separate determination of the unknown parameters with high precision is not available. The values of $E_C$ and $\beta$ used in the calculation are consistent with the values expected for the observed junction areas. Near the calculated curves the corresponding energies are denoted in terms of the excitation energies $E_{0j} = E_j - E_0$, where $E_0$ is the energy level of the ground state and $E_j$ ($j = 1, 2, 3$) are those for excited states.

Although we do not find quantitative agreement, the origin of the resonances can be inferred reliably on the basis of this calculation. It is remarkable that the one- or two-photon resonances associated with the excitation energy $E_{02}$ from the ground state to the second excited state is observed for all the values of $I_{cont}$ used in the measurement. As expected, the branches for $E_{01}/2$ and $E_{02}/2$ exhibit avoided crossings [Fig.

3(a)], whereas those for $E_{01}$ and $E_{02}/2$ simply cross each other [Figs. 3(a)–(c)]. The comparison between Figs. 3(b) and 3(e) implies that, in Fig. 3(b), the faint branch just below that corresponding to $E_{02}$ can be most likely attributed to the energy of $E_{03}/2$, which is the two-photon resonance for excitation from the ground state to the third excited state. Figure 4 shows the variation in $I_{SW}$ as a function of $I_{mag}$ for $F = 20$ GHz and $I_{cont} = 0$ μA, which corresponds to the topmost region of Fig. 3(b). The resonances attributed to $E_{01}$, $E_{02}$, and $E_{03}/2$ are clearly visible as peaks and dips in this figure.

We note that $E_{02}/2$ is higher than $E_{01}$ near the optimal point, showing the inverted anharmonicity with $E_{12} > E_{01}$.[13] At the operating point, where the branches for $E_{01}$ and $E_{02}/2$ cross each other, the two transition energies $E_{12}$ and $E_{01}$ are degenerate. This operating point may allow novel quantum-information applications, such as qutrit operations, that exploit this degeneracy, but such applications have not been given much consideration.[13]

We now denote by $\Delta_i$ the minimum of $E_{0i}$ ($i$=1, 2, 3) as a function of $I_{mag}$. Because $I_{mag}$ was swept in the measurement, the peak and dip very near the optimal point are not visible, in principle. Nevertheless, we are able to estimate $\Delta_i$ from the visible parts of the energy dispersion curves. The experimental data and the calculated results for $\Delta_i$ are plotted in Fig. 5 for various values of $I_{cont}$. The experimental data are again in qualitative agreement with the calculated result. We note that $\Delta_1$, which is usually referred to as the energy gap, is tuned from 0 to 16 GHz. To the best of our knowledge, this tunable range of the gap is the largest ever reported for a similar 4-JJ flux qubit. Even though the working range of our microwave source was limited to 20 GHz, the observation of the two-photon resonances enabled us to characterize $\Delta_i$ beyond 20 GHz. The energy levels characterized in the present study are consistent with the calculated results from refs. 11 and 12. These levels may be used for novel applications, including those proposed in these references. The adiabatic condition for changing flux,[11,12] which is necessary for implementing these applications, may be easily satisfied in future experiments on this device.

Figure 5 shows that $\Delta_2$ is minimum for $I_{cont} = 5$ μA. This means that $E_{02}$ is minimal at this particular point for variations in $I_{cont}$ and $I_{mag}$. In other words, at this operating point $E_{02}$ is constant to first order for small variations in the fluxes $f_1$ and $f_2$. Therefore, the second excited state at this point might be associated with a relatively long coherence time. This operating point can be viewed as doubly optimal. It should be noted that the transition from the ground state to the second excited state is forbidden at the symmetry point for the 3-JJ flux qubit because of wave-function symmetry.[10] The transition operator for the 4-JJ flux qubit is different from that for the 3-JJ qubit because the flux modulation is effective for both $f_1$ and $f_2$. Therefore, this transition may be allowed for the 4-JJ flux qubit, even at the symmetry point.

Although the calculation described herein is a simplified calculation in which some factors were neglected, we believe that this calculation suffices to find the origin of the observed resonances. To find better agreement between the observed energy spectra and the calculated spectra, we believe it is necessary to take into account the inductance effect associated with the qubit loops (main loop and α-loop) and the flux shift due to the current in the SQUID loop. Another point that was neglected in the present calculation and that may change the result is the asymmetry in the α-loop. However, even upon considering the α-loop asymmetry in the Hamiltonian used for the calculation, the agreement between the observed result and the calculated result did not improve.

The inductance effect has been investigated theoretically by some researchers in the case of a 3-JJ flux qubit for a limited range of α.[17-20] When one takes the inductance effect into account, the commonly used two-phase Hamiltonian is not appropriate. Instead, a three-phase (or four-phase) Hamiltonian should be used for a 3-JJ (or 4-JJ) flux qubit; however, this would make the calculation very tedious. It has been shown that for a 3-JJ qubit, the inductance effect renormalizes the Josephson coupling energies used in the effective two-phase theory.[18] For the sample under investigation, the loop length of the main loop was 52 μm. The dimensionless inductance[18] $2\pi L I_c/\Phi_0$ was estimated to be approximately 0.2, where $L$ is the loop inductance, which is dominated by the kinetic inductance for this sample. Therefore, the inductance should have an appreciable effect on energy levels. We should view the values of the parameters used in the present calculation based on the effective two-phase Hamiltonian as renormalized parameters. Theoretical studies have shown that the inductance has an appreciable effect on the energy gap between the two lowest levels in a 3-JJ flux qubit.[17-20] It effectively increases the barrier height between the potential minima, resulting in a decreased tunneling amplitude and energy gap.

For the sample we studied, in addition to the inductance effect of the main loop, that of the α-loop might have an observable effect because its size is comparable to that of the main loop (see Fig. 1). The inductance effect associated with the α-loop has not been studied. The flux shift due to the SQUID bias current[21] must also have an observable effect on the energy dispersion curves, because our data were obtained from the switching-current measurement with a relatively slow sweep of the SQUID bias current. The flux bias in the qubit loop depends significantly on the SQUID bias current. It is difficult to estimate this flux shift quantitatively because the DC-SQUID in this particular sample probably has very asymmetric critical currents of which the values are hard to determine because of the partially broken tunnel barriers. A full understanding of these effects on the energy spectrum requires further study.

In order to observe resonances at an optimal point, it is common to take spectroscopic data with sweeping microwave frequency at a fixed magnetic flux. However, we were not able to identify the qubit resonances by sweeping frequency because the dependence of $I_{SW}$ on frequency exhibited a large number of resonances due to environmental degrees of freedom. A possible source of these resonances could be the large sample box we used, of dimensions of 23 mm × 15 mm × 4mm. It is noteworthy that even in the presence of many extrinsic resonances, we can observe a clear energy dispersion curve of a qubit by using a simple technique involving continuous microwaves and sweeping magnetic flux. Our method uses a large sample box and a loop antenna and is particularly useful for examining several samples in a single experimental run. We also carried out a spectroscopic measurement by recording the switching probability in a pulsed scheme.[21,22] With sweeping $I_{mag}$, we then obtained similar spectroscopic data to those taken by the switching current measurement. Rabi oscillations have not been observed for this sample. The presence of many extrinsic resonances and external noise associated with the sizable control line coupled to the α-loop might be the cause for the significant relaxation and dephasing in this measurement. Furthermore, the coupling strength of the loop antenna to the qubit might be too weak to realize measurable Rabi oscillations.

Figure 6 shows the calculated value of $\alpha_{eff}$ [eq. (5)] and the potential barrier height together with the observed energy gap $\Delta_1$. The barrier height is here defined as the energy difference between the two degenerate minima and the saddle point in the potential energy surface. We see that, with decreasing $I_{cont}$, $\Delta$ increases first exponentially and then linearly, whereas $\alpha_{eff}$ decreases almost linearly throughout the entire range of measurement. It is noteworthy that, for $I_{cont} < -15$ μA, $\alpha_{eff}$ is less than 0.5 and the barrier height becomes zero, which indicates that the potential is a single well in this region. Although the boundary between the double-well regime and the single-well regime cannot be determined precisely because of the error associated with the simplified calculation, it is most probable that the potential of the system undergoes a crossover from a double-well to a single-well regime as $I_{cont}$ decreases in this measurement. We note that a single-well regime of a 3-JJ and 4-JJ qubit wherein α < 0.5 has not previously been investigated theoretically or experimentally, except for the very recent report on a 3-JJ qubit.[13] It is not appropriate to refer to the system in this regime as a flux qubit, which is commonly associated with a double-well potential.

For a tunable RF-SQUID[23] and three-loop 3-JJ flux qubit,[24] a single well potential was used to demonstrate coherent oscillations without microwaves. In this scheme, the initial state, prepared in a double-well potential, was converted to a superposition of the ground state and first excited state in a single-well potential. After the double-well potential was restored, the flux state was measured. The intermediate use of the states in the single-well potential has been shown to be very effective to achieve fast coherent oscillations up to 21 GHz due to the large energy interval between the ground state and first excited state.[23]

Our static spectroscopic measurements in a 4-JJ flux qubit imply that it is possible to initialize, manipulate, and readout a quantum state in a fixed single-well potential with the technique demonstrated in ref. 21, which does not require fast manipulation of the potential profile. Specifically, readout by a simple switching-current measurement in the single-well regime is demonstrated for a sample with an experimentally feasible size, because the flux signal in this regime (at least in the crossover region) is still large enough because of the weak α dependence of the circulating current (away from the optimal point). Note that the manipulation of the state at the optimal point where the decoherence is minimal and the readout far from that point are compatible both in the single-well and double-well regimes using the phase-bias shift due to the SQUID bias current[21] or current pulse.[7]

The use of the single-well regime in a 3-JJ and 4-JJ qubit may allow faster operations of a qubit than in the double-well regime because of the energy level separation larger than 10 GHz, similar to ref. 23. In a 3-JJ and 4-JJ qubit, the loop inductance can be very small; therefore, the loop size can be made much smaller than that of a tunable RF-SQUID. A smaller loop size could result in higher immunity to magnetic field fluctuations. This is a great advantage of either a 3-JJ or a 4-JJ qubit over a tunable RF-SQUID, where a significant loop inductance is necessary in order to produce the relevant potential.

Another advantage of the single-well regime over the double-well regime is the lower sensitivity to magnetic flux fluctuations of the energy intervals, which results in an enhanced level of quantum coherence. The decrease in the flux-noise effect on quantum coherence with decreasing α was predicted in ref. 14, although α was limited to be greater than 0.5 in those calculations. Very recently, it has been reported[13] that a 3-JJ flux qubit in the single-well regime with α = 0.43 has a long coherence time of over 1 μs. We note that, in future experiments on the device under investigation, the α dependence of the coherence time could be studied, which would be very interesting.

When one considers qubit dynamics, which are favorably confined in the two lowest levels, leakage to the higher states is an important issue.[25] As shown in Fig. 5, deep in the double-well regime the two-level approximation is very good because $E_{12}$ is much larger than $E_{01}$. As the system enters the single-well regime as a result of decreasing $I_{cont}$, $E_{12}$ gradually approaches $E_{01}$. Therefore, in the single-well regime, leakage to the second excited state may be generally larger than in the double-well regime. For further understanding of this leakage, the transition matrix elements between the two lowest states and the higher states need to be studied.[25]

The study of these transition matrix elements is also crucial for the realization of the proposed applications[11,12] using the three lowest levels. We note that, in calculating the transition matrix elements described in refs. 11 and 12, it is assumed that the microwave-field modulation is only applied to the main loop of the qubit. This is not valid for the experiment presented herein, where the areas of the main loop and the α-loop are comparable. The field modulation for both loops should be taken into account in calculating the transition matrix elements for the sample under investigation.

## 5. Summary

In summary, we performed spectroscopic measurements of a double-loop 4-JJ flux qubit with a tunable energy gap by using a simple technique that uses continuous microwaves and slow sweeps of the SQUID bias current. The gap was tuned from 0 to 16 GHz by varying the applied bias magnetic flux. In addition to the excitation energy from the ground state to the first excited state, that from the ground state to the second excited state was determined as a function of $I_{mag}$ for various values of $I_{cont}$ by observing one- and two-photon resonances. A branch of the spectrum that is most likely attributable to the two-photon resonance to the third excited state was also observed. We experimentally found the doubly optimal point for excitation from the ground state to the second excited state. This operating point might be associated with a long coherence time. The observed excitation energy spectra were in qualitative agreement with the results of simplified two-phase calculations, where the effects of inductances and flux shifts due to the SQUID bias current were neglected. Our demonstration of tunability of the energy levels of the three lowest states in this system is a first step toward the implementation of applications that use these levels, including the recently proposed schemes for single photon production and cooling. We identified the operating point at which the two transition energies $E_{12}$ and $E_{01}$ were degenerate and novel quantum-information applications involving three equidistant levels may be possible.

Calculations showed that the potential of the system underwent a crossover from a double well to a single well when the applied magnetic flux was varied in the measurements. Our result indicated that it is possible to initialize, manipulate, and readout a 3-JJ and 4-JJ qubit in a single-well regime by using a simple technique that does not require fast manipulation of the potential profile. The qubit operation in this regime is potentially faster than in the commonly used double-well regime because of the much larger energy interval between the two lowest levels in the single-well regime. A recent experiment involving a 3-JJ qubit in the single-well regime demonstrated excellent coherence times of over 1 μs. It would be very interesting to investigate the α dependence of coherence time using a sample similar to the one used in our research, i.e., with a tunable α.

**Acknowledgment**

The author thanks Professor F. Nori for valuable comments on the manuscript. This work was supported by a Grant-in-Aid for Scientific Research from the Japan Society for the Promotion of Science.

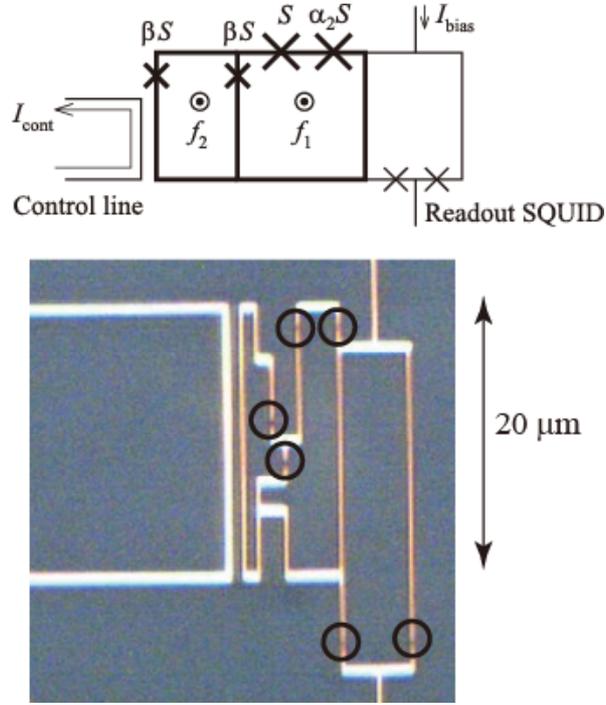

Fig. 1. Schematic representation and optical microscopy image of sample. The 4-JJ qubit is denoted by thick lines in the schematic diagram. The small Josephson junctions are represented by crosses in the circuit diagram and are circled in the photograph. The applied current flowing in the control line modifies the magnetic flux $f_2$, thereby changing the gap. The sample under investigation had $\alpha_2 = 0.8$.

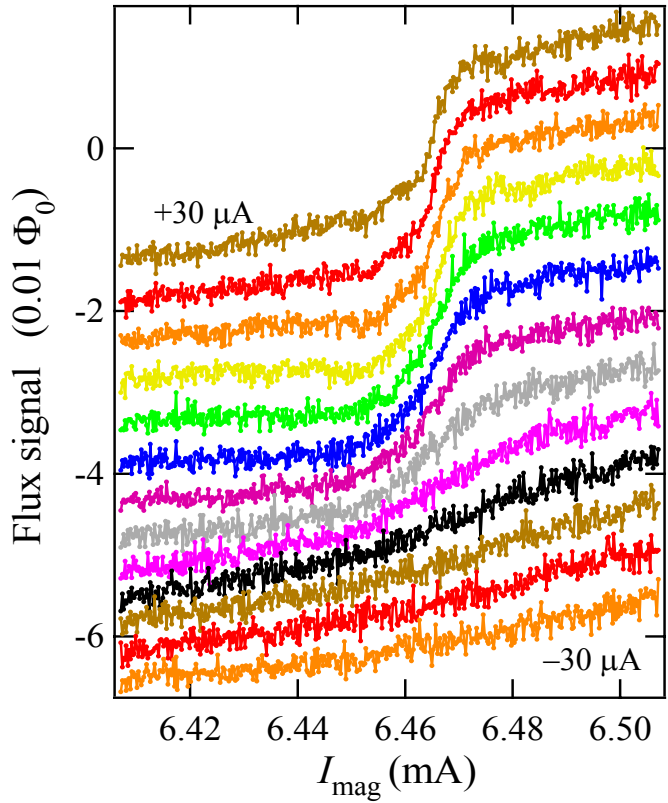

Fig. 2. (color online) Flux signal measured by readout DC-SQUID as a function of $I_{mag}$ for various values of $I_{cont}$. $I_{cont}$ ranges from −30 to 30 μA. For clarity, the curves are offset both vertically and horizontally. The gradual disappearance of the qubit step is apparent. The values of $I_{mag}$ shown here are different from those shown in Figs. 3 and 4 because the data were acquired in different experimental runs.

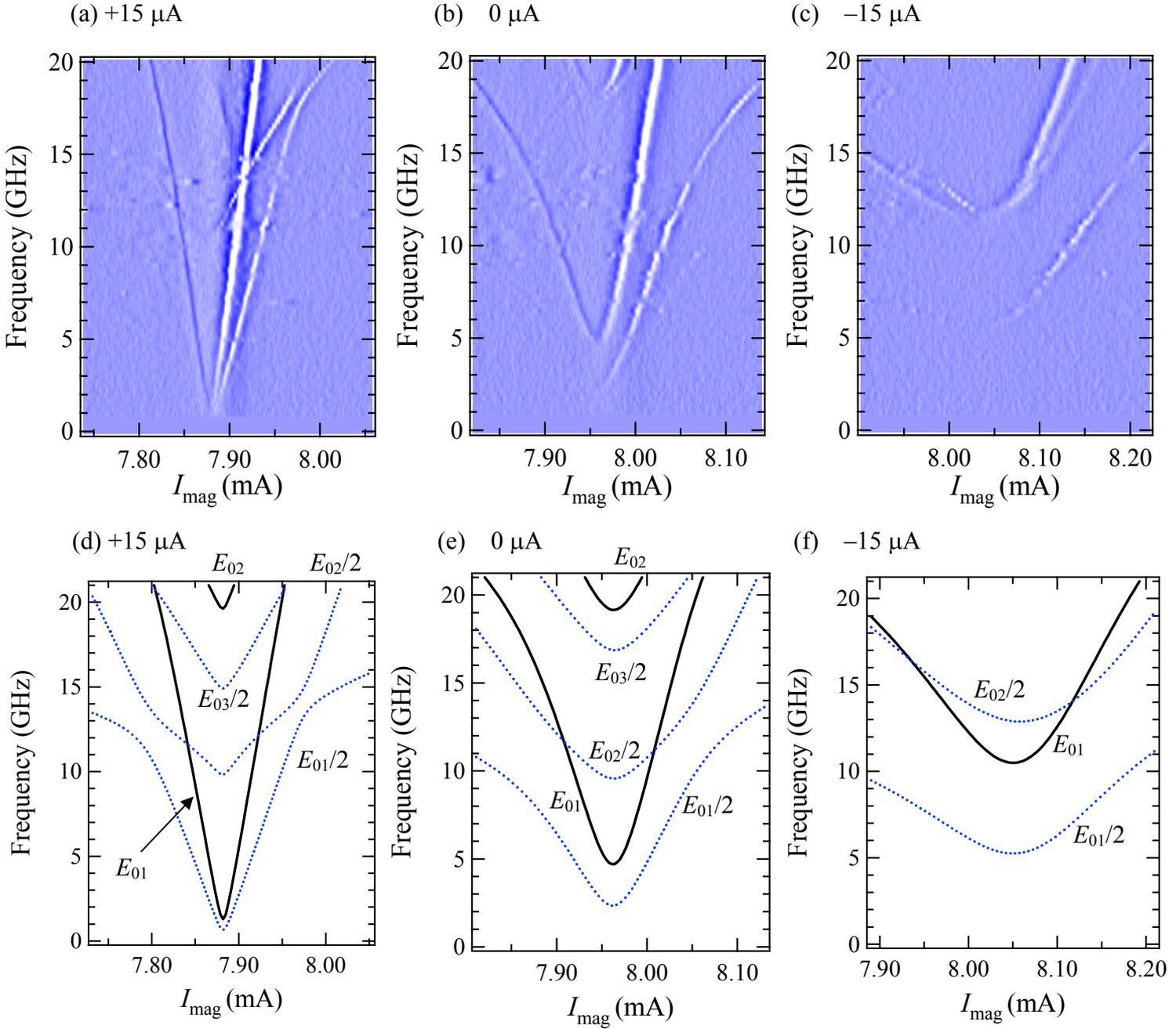

Fig. 3. (color online)
(a)–(c) Grayscale images of spectroscopic data in plane of microwave frequency $F$ and $I_{mag}$ for $I_{cont}$ = 15, 0, and –15 µA. The grayscale indicates peak and dip structures in the $I_{SW}$ vs. $I_{mag}$ curves measured at constant $F$. The dark and light regions represent peaks and dips, respectively. Because $I_{mag}$ was swept in the measurement, the peak or dip structures at the optimal point are not visible. The reason for the asymmetric shape of the energy dispersion curves shown in panels (b) and (c) is that the potential-barrier height depends strongly on $I_{mag}$.
(d)–(f) Calculated excitation-energy spectra showing one- and two-photon resonances. These were calculated using the parameters $E_J$ = 148 GHz, $E_J/E_C$ = 50, β = 0.54, and $f_2^{(0)}$ = –0.325. The corresponding energies are shown near the curves in terms of excitation energies $E_{0j} = E_j - E_0$. The one-photon resonance (solid line) is associated with $E_{0j}$, whereas the two-photon resonance (dotted line) is associated with $E_{0j}/2$.

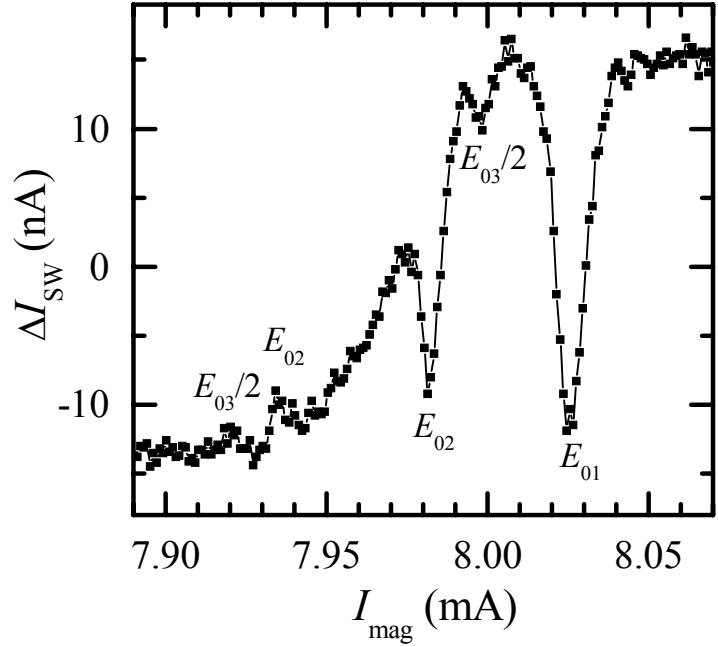

Fig. 4. Variation of switching current $I_{SW}$ of readout DC-SQUID as a function of $I_{mag}$ for $F = 20$ GHz and $I_{cont} = 0$. The background of $I_{SW}$ was subtracted. The resonant peaks and dips are assigned as indicated.

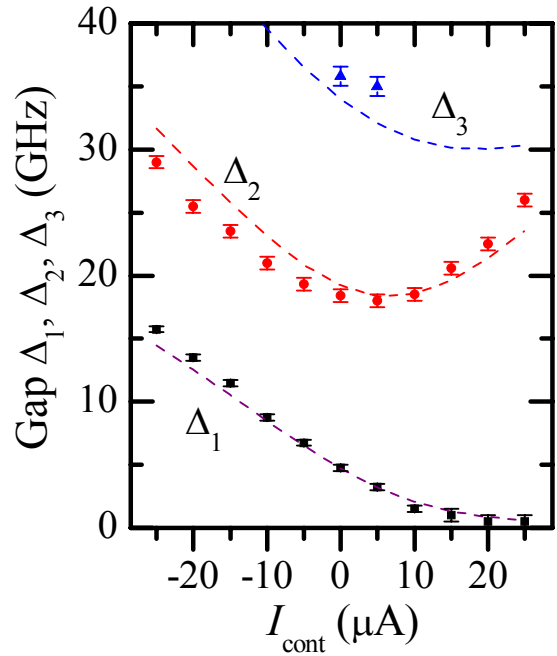

Fig. 5. (color online) Minimum excitation energies $E_{0j}$ (denoted as $\Delta_j$) as a function of $I_{cont}$. Dashed lines show the results of calculations performed using the parameters given in the caption of Fig. 3.

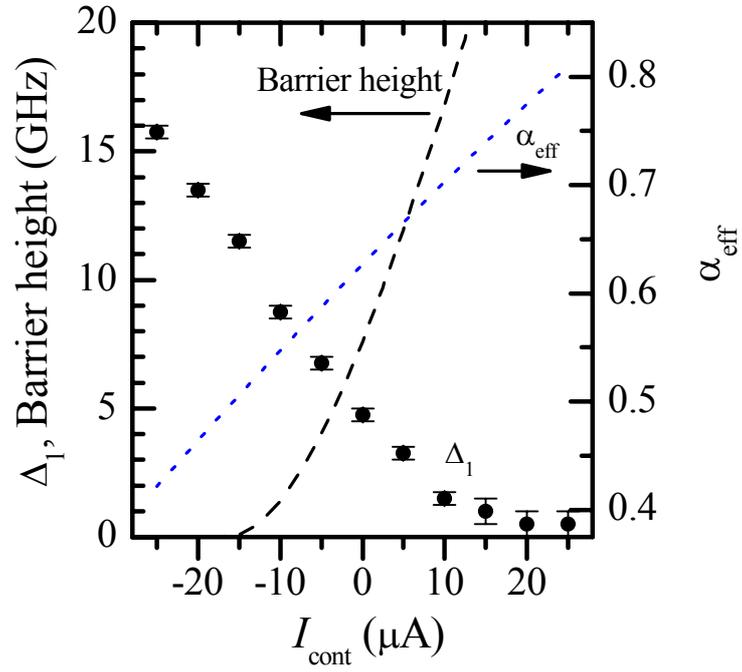

Fig. 6. (color online) Energy gap $\Delta_1$ (solid circles), potential-barrier height (dashed line) at optimal point, and $\alpha_{\mathrm{eff}}$ (dotted line) as a function of $I_{\mathrm{cont}}$. The barrier height and $\alpha_{\mathrm{eff}}$ are calculated using the parameters given in the caption of Fig. 3. In the region of $\alpha_{\mathrm{eff}} < 0.5$, the potential is a single well.